\documentclass{iaus}
\usepackage{graphicx}

\title[Stellar Populations in Spiral Galaxies] 
{Stellar Populations in Spiral Galaxies}

\author[MacArthur \etal]   
{Lauren A. MacArthur$^1$, Jes{\' u}s J. Gonz{\' a}lez$^2$ \and 
 St{\'e}phane Courteau$^3$}

\affiliation{$^1$ California Institute of Technology, Pasadena CA 91125, USA
\break email: lam@astro.caltech.edu\\
[\affilskip]$^2$Instituto de Astronom{\' i}a, UNAM, Mexico\\
[\affilskip]$^3$Queen's University, Kingston, ON K7L 3N6, Canada}

\pubyear{2007}
\volume{241}  
\pagerange{1--4}
\date{?? and in revised form ??}
\jname{Stellar Populations as Building Blocks of Galaxies}
\editors{Peletier, R.F. \& Vazdekis, A., eds.}

\newcommand\apj{ApJ}%
\newcommand\apjs{ApJS}%
\newcommand\araa{ARA\&A}%
\newcommand\mnras{MNRAS}%

\newcommand{\ie}{i.e.\@}

\newcommand\ion[2]{#1$\;${\small\rmfamily{#2}\relax}}
\def\oiii {[\ion{O}{III}]}
\def\nii {[\ion{N}{II}]}
\newcommand{\apgt}{\ {\raise-.5ex\hbox{$\buildrel>\over\sim$}}\ }
\newcommand{\aplt}{\ {\raise-.5ex\hbox{$\buildrel<\over\sim$}}\ }
\newcommand{\apeq}{\ {\raise-.5ex\hbox{$\buildrel<\over=$}}\ }
\newcommand{\gtlt}{\ {\raise-.5ex\hbox{$\buildrel<\over>$}}\ }

\begin{document}

\maketitle

\begin{abstract}
We report preliminary results of the characterization of bulge and inner 
disk stellar populations for 8 nearby spiral galaxies using Gemini/GMOS.
The long-slit spectra extend out to 1--2 disk scale lengths with 
S/N/\AA$>$50.  Two different model fitting techniques, absorption-line indices
and full spectral synthesis, are found to weigh age, metallicity, and
abundance ratios
differently, but with careful attention 
to the data/model matching (resolution \& flux calibration), we are able 
constrain real signatures of age and metallicity gradients in star-forming 
galaxies.

\end{abstract}

\firstsection 
\section{Introduction}\label{intro}

Understanding the formation and evolution of disk galaxies, in
particular the formation of their spheroidal components, remains a
challenge both from the theoretical and observational perspectives.
Traditional views suggest that all bulges formed as ``miniature ellipticals'' 
via dissipationless collapse (\eg Eggen \etal\ 1962) or by the violent 
merging of smaller fragments (\eg Kauffmann 1996) 
at high-redshift.  A robust prediction from these models is that all 
spheroids have old stellar populations (SPs), with bulges being older than 
field ellipticals at a given redshift.  These views
have been challenged by recent observations suggesting ongoing formation
of galactic bulges due, perhaps, to a redistribution of disk material 
from the presence of a bar (Kormendy \& Kennicutt 2004).

Detailed analysis of the SPs in galaxies can provide invaluable clues about
their formation and subsequent evolution.  Broad-band colors have often been 
used as a proxy for SP parameters (\eg MacArthur \etal\ 2004).  
Color-based studies, however, are plagued by a degeneracy between the effects 
of age, metallicity ($Z$), and dust reddening, all leading to redder colors.
Spectroscopic techniques, largely impervious to dust effects (MacArthur 2005), 
offer a more detailed and discriminating view, especially in light of the 
latest implementations of SP synthesis models (\eg Vazdekis 1999; 
Bruzual \& Charlot 2003). 

Existing spectroscopic studies of bulges spanning the full range of Hubble 
types are few, and the results are conflicting.  The absorption-line studies 
of the central regions of Sa\,--\,Sc spirals
of Trager \etal\ (1999) and Proctor \& Sansom (2002, PS02) both find that 
late-type bulges cannot be reproduced using primordial collapse models and 
invoke extended gas infall to explain the observations, while, in a similar 
study, Goudfrooij \etal\ (1999) conclude the opposite.  Recently, 
Thomas \& Davies (2006) reanalyzed the PS02 data and found no difference 
between the SPs of spiral bulges and Es at a given central velocity 
dispersion, $\sigma_0$, concluding that processes involving disk material 
cannot be responsible for the recent star formation implied by the young 
ages.  A significant outlier, however, is the bulge of our own Milky Way (MW), 
studies of which reveal old and $\alpha$-enhanced SPs that must
have formed long ago and on short timescales.  This observation is difficult
to reconcile with the MW's small $\sigma_0$ and the presence of a bar.
As Thomas \& Davies point out, the MW studies sample a larger physical 
radius than those of external galaxies so this conundrum could be resolved 
if there is a positive age gradient in the bulge. 
At the heart of these conflicts is the fact that extant 
studies of local bulges are limited in one way or another (sample size, 
range in Hubble type, data of insufficient depth and/or 
spatial resolution for studies of gradients and assessment of disk 
contamination, etc.)

\section{Program and Observations}

To address the above issues, and to compare different techniques for 
measuring SP parameters for composite systems, we have obtained Gemini/GMOS 
long-slit spectroscopy for a pilot sample of 8 nearby face-on field 
spirals (Sa--Scd) from which we measure a suite of absorption line-indices 
and perform full spectral synthesis fits through the bulges and inner disks.
Our simultaneous spectral coverage of $\sim$4050\,--\,6750\,\AA\ includes 
most of the major atomic and molecular features to disentangle 
age \& $Z$ effects in integrated galaxy spectra.
Integration times of 45\,--\,72 min per galaxy were required to 
achieve adequate S/N/\AA\ ($>$50) into the disk, to constrain disk 
populations and to allow for the removal of disk contamination in the bulge.
Details about the data and reductions, including calibration 
($\lambda$ \& flux), resolution (instrumental \& profiles), rotation,
velocity dispersion, and bulge/disk decontamination will be described 
in Gonz{\' a}lez \etal\ (2007, in prep.).

\section{Stellar Population Indicators}
\subsection{Lick Indices}
As a first attempt at constraining the SPs in our galaxies, we measured
24 spectral (Lick/IDS) indices (missing only the bluest H$\delta_{A}$),
and compared them with the SSP models of BC03.  Many of our galaxies have 
significant line-emission, some extending all the way into the center.  
Many authors attempt to correct for emission by fitting emission-free 
templates to their galaxy spectra (\eg Gonz{\'a}lez 1993).
However, we are interested here in testing the discriminating power of
the indices on their own (\ie\ without the need for accompanying 
higher-resolution spectra).  We thus did not apply any corrections to 
individual indices, but rather used a scheme in which the emission-line 
affected indices are systematically eliminated from the fit of indices 
to SSP models.

By comparing the age \& $Z$ measurements using different index combinations,
depending on the pathologies of the given spectrum, we conclude that the
metallic indicators are fairly reliable but, when emission is present,
there is simply not enough sensitivity in the indices unaffected by 
emission to obtain reliable age estimates.  

\subsection{Spectral Fitting}

We thus turned our attention to full spectral synthesis fits.  While these 
offer the advantage of being able to mask out undesirable regions of the 
spectrum, the wider wavelength-baseline reinforces the sensitivity to dust 
extinction along with fluxing and resolution issues (which must be carefully 
matched between data and models).  Our spectral synthesis technique is an 
optimized linear, bound-constrained (non-zero), combination of BC03 SSP model 
templates (spanning ages 0.001--20Gyr and $Z=(0.02-2.5)Z_{\odot}$), while 
masking out regions affected by emission, persistent sky lines, and CCD gaps.  
Best-fit SSP and mixed populations were obtained for each radial bin 
of each galaxy.
\begin{figure}
\begin{center}
\includegraphics[width=0.90\textwidth]{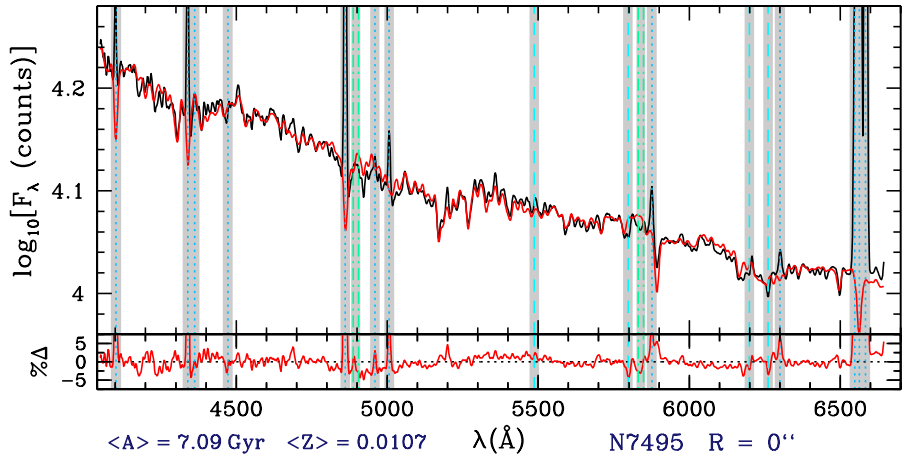}
  \caption{Full spectral synthesis fit to the central spectrum of the
           emission-line dominated Sc galaxy NGC 7495.  Black line: data.  
           Red line: model fit.  Shading indicates emission-line regions
           (including the Balmer series, \oiii, \nii, etc.,
           are indicated by vertical dotted blue lines), 
           variable night sky lines (dashed light blue lines), and
           CCD gaps (dashed light green line), that are excluded from the fit. 
           Residuals are shown as percent galaxy$-$model differences.}
\label{fig:specfit}
\end{center}
\end{figure}
Fig.~\ref{fig:specfit} shows an example of our full synthesis fits for a 
spectrum showing strong emission.  The data-model agreement is quite 
remarkable throughout the entire spectrum.  Analysis 
of the radial bins for each galaxy also demonstrated the stability
of our full spectral fitting technique, whereas the SSP fits were unstable
producing spurious gradients.

\begin{figure}
\begin{center}
\includegraphics[width=0.48\textwidth]{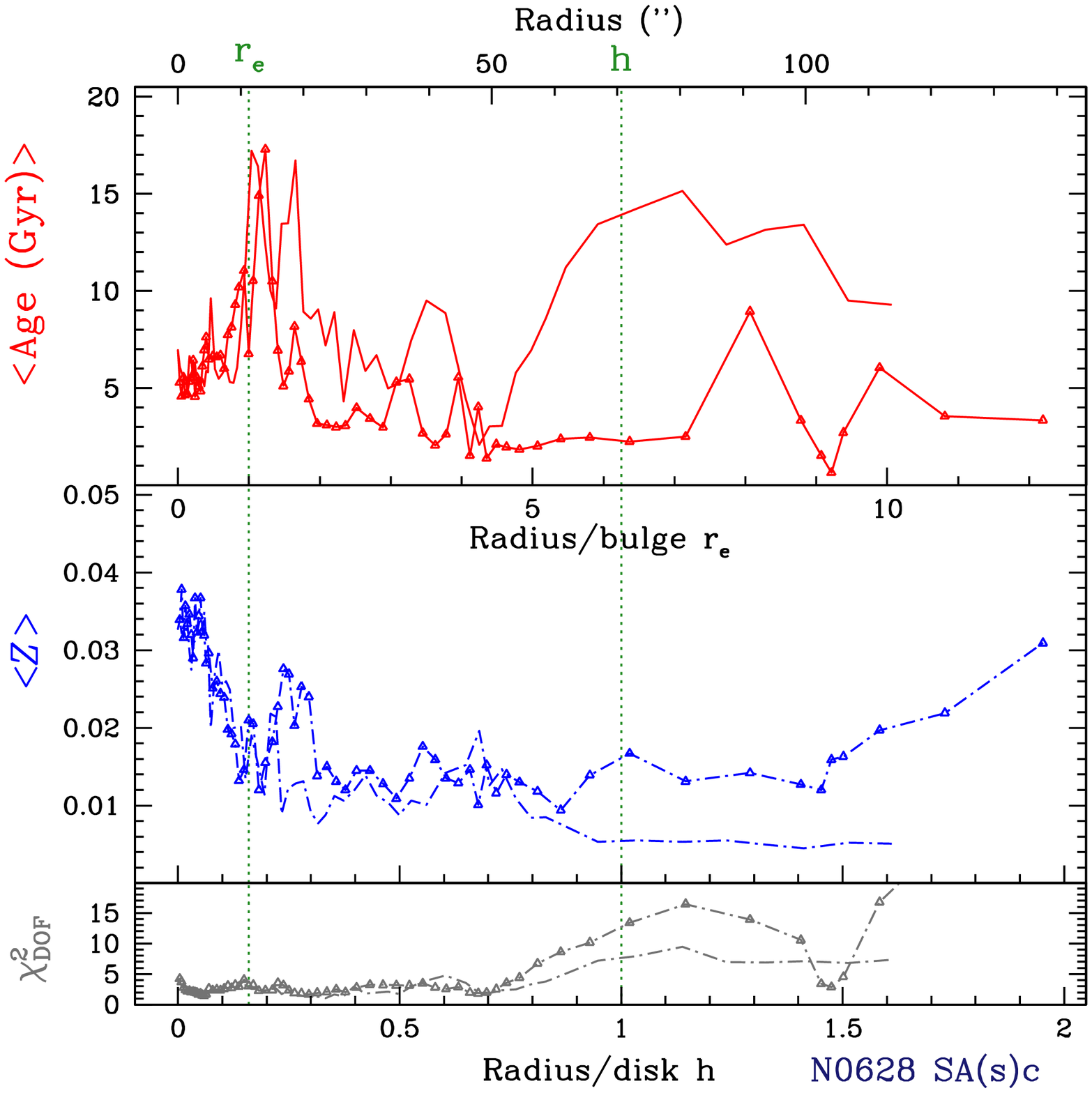}
\includegraphics[width=0.48\textwidth]{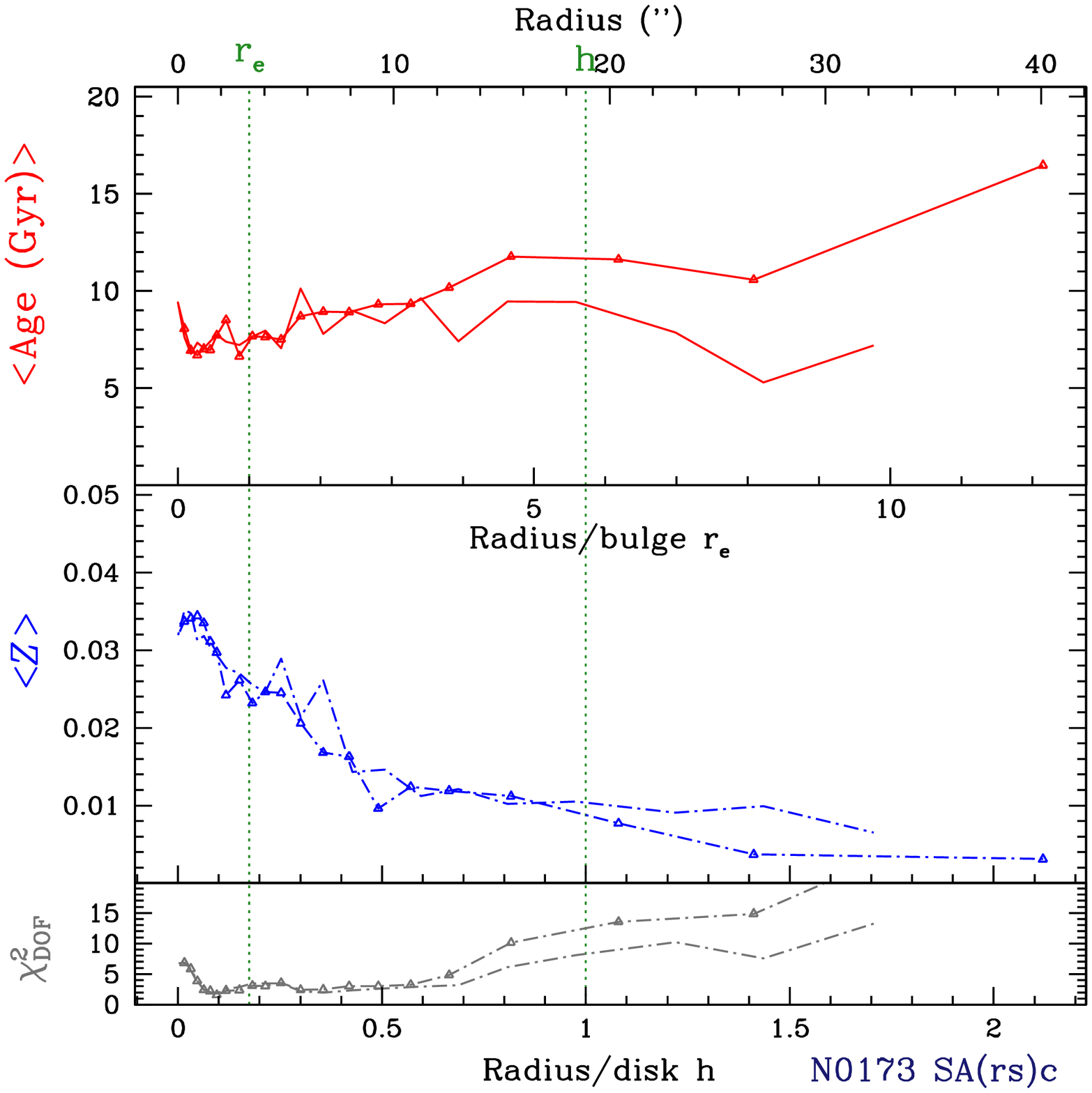}
  \caption{Light-weighted gradients in average age (top panels) and 
           metallicity (middle panels) from full spectral synthesis fits.  
           The reduced $\chi^2$ from the fit is shown in the bottom panels.
           The two distributions in each panel indicate the two sides of the 
           galaxy with the common side marked by triangles.  Vertical dotted
           green lines mark the bulge effective (half-light) radius, $r_e$, 
           and 1 disk scale length, $h$.}
\label{fig:AZgrads}
\end{center}
\end{figure}

Our study is the first to provide radially resolved spectra well into 
galaxy disks, enabling a comparison of the bulge and inner disk populations.
Fig.~\ref{fig:AZgrads} shows two examples of the radial profiles of 
average age \& $Z$ as determined from our full spectral synthesis fits. 
Of particular note is the spike to older ages just beyond $\sim$1$r_e$
for NGC 628.  This is precisely the signature noted in \S\ref{intro}
required to reconcile the discrepant results for the MW bulge.  The
fact that we observe this signature on both sides of the galaxy (whose 
spectra were fit entirely independently) renders confidence that this 
signature is real.\footnote{Using SAURON observations, Gandia {\it et al.} 
(in prep.) observe the same trend in age for NGC 628 using a fully 
independent approach.}
While the small size of our sample does not allow for a detailed statistical
study of galaxy parameters with inferred age \& $Z$, several important
observations emerge: (1) late-type bulges show recent star formation in 
their centers, potentially driven by secular evolution processes; (2) most 
bulges show positive age gradients (younger toward the center) and negative 
$Z$ gradients; and (3) inner disks show weak age gradients (if any) 
and all have significant negative or flat radial $Z$ profiles.
We must caution that to truly assert the reality of young bulge SPs requires
the careful removal of any contamination of disk light.  This is indeed 
a major focus of our continued investigation of this rich data set.  

\section{Summary}\label{sec:conc}
Using radially resolved long-slit spectra of 8 star-forming galaxies, we 
have assessed the strengths and weaknesses of different SP analysis 
techniques in the presence of emission lines.  The most pertinent 
observations are as follows:

\begin{itemize}
\item With moderate spectral resolution, good $\lambda$ coverage, and
      high S/N/\AA\ ($>$50), measurement of light-weighted ages \& 
      metallicities for star-forming galaxies is feasible.
\item Details are critical: calibration ($\lambda$ \& relative flux),
      resolution, velocity dispersion, and rotation must be treated
      self-consistently between the data and models.
\item Different fitting techniques weigh age, metallicity, and 
      abundance ratios differently:  Balmer emission limits age
      fitting from indices; age information is recovered (in the
      presence of emission) from full spectrum and continuum SED
      fitting (but compounds the caveats about dust extinction and 
      fluxing that are less important for indices).
\item SSPs are not a good match to late-type galaxies.  The degeneracies
      between age, metallicity, dust, etc. are extreme leading to 
      unstable fits.
\item Full spectral synthesis fitting is consistent with metallicity 
      measured via indices, but is far less sensitive to abundance 
      variations.
\end{itemize}

Ultimately, we desire a large enough sample to look for trends with galaxy 
parameters allowing for the discrimination between different scenarios for 
current models of galaxy formation and to guide future implementations of 
such models.

\begin{acknowledgments}
We would like to thank the conference organizers for putting together
such a stimulating and informative meeting.
\end{acknowledgments}





\end{document}